\documentclass[twocolumn,aps,prd,epsf]{revtex4-1}%
\usepackage[dvips]{graphicx}
\usepackage{epsfig}
\usepackage{color}
\usepackage{amsmath}
\usepackage{amsfonts}
\usepackage{amssymb}%
\newcommand{\be}{\begin{equation}}
\newcommand{\ee}{\end{equation}}
\newcommand{\bea}{\begin{eqnarray}}
\newcommand{\eea}{\end{eqnarray}}
\newcommand{\non}{\nonumber}

\begin{document}
\title{Computation of pion and kaon heavy ion multiplicities in a gluon-meson model}
\author{Pedro Bicudo}
\email{bicudo@ist.utl.pt}
\affiliation{CFTP, Dep. F\'{\i}sica, Instituto Superior T\'ecnico, Av. Rovisco Pais,
1049-001 Lisboa, Portugal}
\author{Francesco Giacosa}
\email{giacosa@th.physik.uni-frankfurt.de}
\affiliation{Institute of Theoretical Physics, J. W. Goethe University, Max-von-Laue Str.
1, D-60438, Frankfurt am Main, Germany}
\author{Elina Seel}
\email{seel@th.physik.uni-frankfurt.de}
\affiliation{Institute of Theoretical Physics, J. W. Goethe University, Max-von-Laue Str.
1, D-60438, Frankfurt am Main, Germany}

\begin{abstract}
In high energy Heavy Ion Collisions, the onset of the quark-gluon plasma is
the colour glass condensate, dominated by gluons. The final state is hadronic,
and dominated by pions and kaons. Here we investigate an effective approach of
QCD with these bosonic fields and which can help to describe the transition of
gluons into light mesons. Formally, our approach consists in
integrating out the quark fields from the QCD path integral. In this way the fermionic fields are replaced by light
mesons, such as the pions and sigma field. We apply our effective action to compute the number
of pions and kaons per gluon emitted by a Boltzmann gluon gas, their
multiplicities as a function of the gluon mass. We conclude that an effective gluon mass remains finite at $T=T_c$.
\end{abstract}
\maketitle

%S
%S
%S
%S
%S
%S
%S
%S
%S
%S
%S
%S

\section{Introduction}

In this work we develop a Lagrangian where only bosons, i. e. gluons and mesons, are the
active degrees of freedom, and apply it to study the multiplicities in heavy ion
collisions, the number of pions and kaons per gluon, and the gluon mass at the
onset of the deconfinement/confinement phase transition
\cite{dirkrev}.

Effective approaches of QCD have been widely used to study the properties of strong
interactions \cite{amslerrev}. Quark models, meson effective models, or models combining both
quarks with mesons are used thoroughly to explore hadronic physics. Although gluons have
been proposed already in the 70's together with the theory of strong
interactions, QCD, in effective models it is common to
assume that the gluons are integrated out, and only contribute indirectly 
through the quark or effective hadron interactions.

Nevertheless, there are two rapidly developing QCD domains where gluons are
either easier to work with, or are phenomenologically more relevant. 
In many-body systems, the Grassmann variable nature of the quarks makes them
technically much more difficult to address than the bosonic gluons. In
particular, Lattice QCD first developed and applied pure gauge or quenched
techniques since working with dynamical quarks is computationally very
expensive \cite{Gattringer:2010zz,Creutz:1998ee}. Moreover, in high energy Heavy Ion
Collisions \cite{dirkrev}, it was proposed successfully that the onset of the quark-gluon plasma is the colour glass condensate, dominated by gluons 
\cite{McLerran:1993ni,JalilianMarian:1997jx,Iancu:2001ad,Ferreiro:2001qy}. 
The final state is hadronic, and dominated by
pions and kaons. For instance, in the many particle BAMPS set-up for heavy ion collisions,
\cite{Xu:2008av,Bouras:2009nn} the simulations are performed with gluons only
form the onset, and mesons are included as final states of the hadronization.
From the QCD perspective, in this case, quarks and not gluons are integrated
out and effective mesons are included.
Note also the recent work of Weinberg \cite{weinberg} where 
an effective Lagrangian with gluons, in addition to pions and quarks, was put forward.

Formally, the different effective approaches to QCD can be seen
as the result of integrating out some degrees of freedom from the QCD
Lagrangian \cite{amslerrev}. For instance, when only gluons are integrated out,
one obtains a NJL-like theory \cite{njl} or a quark model \cite{De Rujula:1975ge}.
If, in addition, also quarks are integrated out
from the NJL model, one is left with a purely linear $\sigma$ model. Moreover, as an intermediate step between the NJL and the
$\sigma$ models a quark-meson model is obtained. Similarly, in the approach of
Cahill and Roberts \cite{Cahill:1985mh,Praschifka:1986nf} it was shown by using bilocal
auxiliary fields (along the same line of the Hubbard-Stratonovich
transformation \cite{Hubbard:1959}) how to integrate out quark and gluon
degrees of freedom to obtain a purely mesonic Lagrangian. While these
calculations were only performed at one loop order, it was an interesting
approach to connect effective models, say the $\sigma$ model, directly to QCD.
In addition, there are also lattice QCD approaches for effective meson
theories \cite{Canosa:1994su}, which deliver qualitatively similar results.

For the purpose of this paper it is necessary to chose a slightly different
way, which consists of integrating out from the QCD Lagrangian the quarks only
and then obtain a gluon-meson (fully bosonic) theory. From symmetry principles
(colour gauge invariance and chiral symmetry) we expect at leading order the
following tree-level coupling between the gluons, the pions and their chiral
partner, the scalar $\sigma$ meson:
\begin{equation}
\mathcal{L}_{\text{gluon-meson}}\propto(G^{a,\mu\nu}G_{\mu\nu}^{a})(\sigma
^{2}+{ \vec \pi}^{2}+...)\ , \label{simple}%
\end{equation}
where $G_{\mu\nu}^{a}$ is the gluonic field tensor and dots
refer to other mesonic degrees, such as resonances with open and hidden
strange-quark content (such as the kaons), vector resonances and, eventually,
non-quarkonium resonances. 

The interaction Lagrangian (\ref{simple}) allows to
study the transition of two gluons into a couple of mesons. If the latter are
not stable, they further decay into pions or kaons. For the mesonic sector we apply the most comprehensive $\sigma$ model, 
including not only scalar and pseudoscalar mesons, 
but also vector and axial-vector mesons 
 \cite{geffen,denis}.
By assuming a thermal bath
for gluons at the hadronization point, we can then evaluate how many pions
(and kaons) are obtained per gluon as function of the temperature $T$ of the bath.
To account for a possible finite scale with an energy dimension, in the gluon sector at finite $T$, we allow for a finite effective gluon mass in the Boltzmann distribution.

The paper is organized as follows: in Sec. II we derive our effective
Lagrangian, in Sec. III we present the analytical and numerical results, and
in Sec. IV we derive our conclusions.

%S
%S
%S
%S
%S
%S
%S
%S
%S
%S
%S
%S

\section{The gluon-meson interaction}

In this section we present the formal steps necessary to obtain a gluon-meson
theory. We start from the euclidean QCD generating functional%

\begin{align}
Z\left[  \bar{\eta},\eta\right]   &  =\int\mathcal{D}\bar{q}\mathcal{D}%
q\mathcal{D}A \mathcal{D}\bar{\omega}\mathcal{D}\omega\label{1}\\
&  \times e^{-\int_{x}\mathcal{L}_{QCD}+\mathcal{L}_{GF}+\mathcal{L}%
_{FPG}-\bar{\eta}q-\bar{q}\eta}\text{ ,}\nonumber
\end{align} 
where%
\begin{align}
\mathcal{L}_{QCD}  &  =\bar{q}\left( \gamma_{\mu}D^{\mu}+m\right)
q+\frac{1}{4}G_{\mu\upsilon}^{a}G_{a}^{\mu\upsilon}\text{ ,}\\ \non
D_{\mu}  &  =\partial_{\mu}-iA_{\mu}\text{ , \ \ \ }A_{\mu}=gA_{\mu}^{a}%
\frac{\lambda^{a}}{2}\text{ .}%
\end{align}
$\bar{\eta}$ and $\eta$ are the fermion sources, $\mathcal{L}_{GF}$ and
$\mathcal{L}_{FPG} $
represent the gauge-fixing and the Faddeev-Popov ghost terms. 
In a suitable gauge 
\cite{Cahill:1985mh,Praschifka:1986nf}, and with no fermion sources, the generating functional of QCD can be written as
\begin{equation}
Z=\int\mathcal{D}\bar{q}\mathcal{D}q\mathcal{D}A \, e^{-\int_{x}\bar
{q}\left( \gamma_{\mu}D^{\mu}+m\right)  q
- \frac{1}{2} \int_{x} A_{\mu}^{a} {({D^{-1})}_{\mu\nu}^{ab}} A_{\nu}^{b}}
\text{ ,} \label{zqcd}%
\end{equation}
where
\begin{align}
D_{\mu\nu}^{ab}(x-y)  &  =\int\mathcal{D}\bar{\omega}\mathcal{D}%
\omega\mathcal{D}A \, A_{\mu}^{a}(x)A_{\upsilon}^{b}(y)\\ \non
&  \times e^{-\int_{x}\frac{1}{4}G_{\mu\upsilon}^{a}G_{a}^{\mu\upsilon
}+\mathcal{L}_{GF}+\mathcal{L}_{FPG}}%
\end{align}
is the exact gluon propagator which contains all gluon self-interactions and
gluon-ghost interactions but excludes quark loops.

To introduce the mesons we multiply Eq. (\ref{zqcd}) by
\be
1={ 1 \over \cal N} \int\mathcal{D}\Phi\mathcal{D}R_{\mu}\mathcal{D}L_{\mu}e^{-\int
_{x}\mathcal{L}_{\text{meson}}}\text{, \ \ }\label{mesonlag}
\ee
where
\begin{align}
\mathcal{L}_{\text{meson}}  &  =Tr\left[  \partial^{\mu}\Phi^{\dagger
}\partial_{\mu}\Phi\right]  -m_{0}^{2}Tr\left[  \Phi^{\dagger}\Phi\right] 
\\
&  -\frac{1}{4}\mathrm{Tr}\left[  R_{\mu\nu}^{2}+L_{\mu\nu}^{2}\right]
+\frac{m_{1}^{2}}{2}\mathrm{Tr}\left[  R_{\mu}^{2}+L_{\nu}^{2}\right]  \nonumber
\end{align}
is the quadratic part of the globally
invariant linear $\sigma$ model Lagrangian with $U(N_{f})_{R}\times U(N_{f})_{L}$
symmetry: $\Phi,$ $L_{\mu}$ and $R_{\mu}$ are $N_{f}\times N_{f}$ Hermitian
matrices for the (pseudo)scalar, vectorial right-handed and left-handed
mesonic degrees of freedom \cite{denis}:
\begin{align}
\Phi &  =S^{a}t^{a}+iP^{a}t^{a}\text{ ,}\\ \non
L_{\mu}  &  =L_{\mu}^{a}t^{a} = V_\mu + A_\mu \text{ ,} \\ \non
R_{\mu } & =R_{\mu}^{a}t^{a} = V_\mu - A_\mu \text{ ,}\\ \non
R^{\mu\nu}  &  =\partial^{\mu}R^{\nu}-\partial^{\nu}R^{\mu},L^{\mu\nu
}=\partial^{\mu}L^{\nu}-\partial^{\nu}L^{\mu}\text{ .}%
\end{align}
In the previous expressions $t^{a}$ are the $N_{f}^{2}$ generators of
$U(N_{f}),$ $S^{a}=\sqrt{2}\bar{q}t^{a}q$ are the scalar, $P^{a}=\sqrt{2}%
\bar{q}i\gamma^{5}t^{a}q$ are the pseudoscalar degrees of freedom, $V_{\mu
}^{a}= \sqrt{2} \bar q \gamma_\mu t^a q $, $A_{\mu}^{a}= \sqrt{2} \bar q \gamma_\mu \gamma^5 t^a q$ are the vector and axial-vector
microscopic quark currents. For instance, in the case $N_{f}=2$ the fields are
given by%
\be
\Phi=\left(  \sigma+i\eta_{N}\right)  t^{0}+(\vec{a}_{0}+i\vec{\pi})\vec
{t}\text{ ,}%
\ee
where $t^{0}=I/2$, $t^{i}=\tau_{i}/2,$ and $\tau_{i}$ are the
Pauli matrices. Analogously, $\ R^{\mu}=\left(  \omega^{\mu}-f_{1}^{\mu
}\right)  t^{0}+(\vec{\rho}^{\mu}-\vec{a}_{1}^{\mu})\vec{t}$ represents the
vector and $L^{\mu}=\left(  \omega^{\mu}+f_{1}^{\mu}\right)  t^{0}+(\vec{\rho
}^{\mu}+\vec{a}_{1}^{\mu})\vec{t}$ \ the axial vector degrees of freedom. The
extension to the case $N_{f}=3$ is straightforward \cite{nf3}.

Clearly, upon Gaussian integration, Eq. (\ref{mesonlag}) gives a constant
factor and thus does not change the path integral in the generating functional
Eq. (\ref{zqcd}).

To couple the mesons to fermions, the mesonic fields can be added as
parallel transports in the fermion matrix similar to a mass term in the case
of the scalar sigma and to a chiral transport in the case of the pion. The
corresponding minimal coupling is given by
\begin{equation}
\gamma_{\mu}D^{\mu}\rightarrow\tilde{D}=\gamma_{\mu}D^{\mu}-c_{1}\Phi
-c_{2}\left(  \gamma_{\mu}V^{\mu}+\gamma_{\mu}\gamma^5 A^\mu\right)  \text{ ,}
\label{coupling}%
\end{equation}
where $c_{1}$ and $c_{2}$ are free parameters, and $A^\mu$ is the axial current. 
Finally, performing the Grassmann integration over the fermion fields, we obtain 
a purely bosonic theory
\begin{align*}
Z  &  =\int\mathcal{D}\Phi\mathcal{D}L_{\mu}\mathcal{D}R_{\mu}\mathcal{D}A
\\
&  e^{-\int_{x}
\frac{1}{2}
A_\mu^a 
{({D^{-1})}_{\mu\nu}^{ab}}
A_\nu^b
+\mathcal{L}_{\text{meson}}}\det\big[  \tilde{D}\big]  
\text{ .}%
\end{align*}
%(removed \frac{1}{4}G^{a,\mu\nu}G_{\mu\nu}^{a} )
The coupling of the mesonic and gluonic degrees of freedom resides in the
fermion determinant $\det \big[  \tilde{D}\big]  ,$ which cannot be computed
analytically. However, by requiring local colour gauge invariance and chiral
symmetry and restricting to lowest dimensionality of the interaction
Lagrangian, we are left to the following Lagrangian,
\begin{equation}
\mathcal{L}_{\text{gluon-meson}}=G^{a,\mu\nu}G_{\mu\nu}^{a}
\mathrm{Tr}\left[ a
\Phi^{\dagger}\Phi+b\left(  V_{\delta}^{2}+A_{\delta}^{2}\right)  \right] ,
\label{lmid}% 
\end{equation}
where $a$ and $b$ are couplings with dimension Energy$^{-2}$ and describe the
transition from two gluons to two mesons. To obtain the relation between
the parameters $a$, $b$ and the parameters $c_{1}$, $c_{2}$ introduced in Eq.
(\ref{coupling}) it would be necessary to evaluate the fermion determinant
$\det\big[  \tilde{D}\big]  $ analytically. Although it is not possible to
compute the fermion determinant exactly, it is natural to expect that $a\sim
b.$ In the following we will work with the simplified assumption $a=b.$ Note
also that the Lagrangian (\ref{lmid}) makes use of the so-called flavour
blindness of the gluon fields.

\begin{table}[t!]
\begin{center}%
\begin{tabular}
[c]{c|ccccc}\hline
meson & ${M_{m}}$ & $g$ & $N_{\pi}$ & $N_{K}$ & f\\\hline
$\pi$ & 138 & 3 & 1 & 0 & a\\
$\eta$ & 549 & 1 & 3 & 0 & a\\
$\eta^{\prime}$ & 958 & 1 & 3 & 0 & a\\
$K$ & 495 & 4 & 0 & 1 & a\\
$\rho$ & 775 & 9 & 2 & 0 & b\\
$\omega$ & 782 & 3 & 3 & 0 & b\\
$\phi$ & 1020 & 3 & 0 & 2 & b\\
$K^{\ast}$ & 892 & 12 & 1 & 1 & b\\
$a_{0}$ & 985 & 3 & 4 & 0 & c\\
$\sigma$ & 600 & 1 & 2 & 0 & c\\
$f_{0}$ & 980 & 1 & 2 & 0 & c\\
$\kappa$ & 800 & 4 & 1 & 1 & c\\
$a_{0}$ & 1450 & 3 & 4 & 0 & a\\
$f_{0}$ & 1370 & 1 & 4 & 0 & a\\
$f_{0}$ & 1710 & 1 & 0 & 2 & a\\
$\kappa$ & 1430 & 4 & 1 & 1 & a\\
$a_1$ & 1230 & 9 & 3 & 0 & b \\
$f_1$ & 1282 & 3 & 4 & 0 & b \\
$f_1$ & 1420 & 3 & 1 & 2 & b \\
$K_1$ &  1272 & 12 & 1 & 1 & b \\  
\hline
\end{tabular}
\end{center}
\caption{The parameters for each meson pair initially produced and then
decaying into pions and kaons are the mass, degeneracy, number of pions
produced by the meson, number of kaons produced by the meson, the family
factor. For meson octets and their respective chiral partners we assume the same family factor. }%
\label{mesons}%
\end{table}%

In accordance with Parganlija et al.  \cite{denis}, the scalar partners of the pions are
identified with the scalar resonances above $1$ GeV. However, for
completeness, one should also include the scalars below $1$ GeV, which
according to many recent and less recent studies are non-quarkonium states. In
fact, these states can be enhancements in the two-pseudoscalar channel or
tetraquark states, see Refs.  \cite{amslerrev} and references therein for a more detailed
discussion of this point. Here they are coupled to the gluons with an
independent coupling $c$. We shall here test two choices: $c=a,$ i.e. with
equal strength as the other channels, and $c=0,$ where the light scalars are switched off.

The explicit evaluation of the traces delivers
\begin{align}
\mathcal{L}_{\text{gluon-meson}}  & =aG^{a,\mu\nu}G_{\mu\nu}^{a}({\vec{\pi}%
}^{2}+\cdots+{{\vec{a}}_{0}}^{2}+\cdots)\nonumber\\
& +bG^{a,\mu\nu}G_{\mu\nu}^{a}(
{{\vec{\rho} \, }_{\delta}} \cdot {{\vec{\rho} \ }^{\delta}}
+ \cdots + 
{{\vec{a_1}}_{\delta}} \cdot {{\vec{a_1}}^{\delta}}
+ \cdots)\nonumber\\
& +cG^{a,\mu\nu}G_{\mu\nu}^{a}(\sigma^{2}+\cdots)\ ,
\label{lfin}%
\end{align}
where dots refer to further quadratic mesonic interactions listed in Table \ref{mesons}.

Even if higher dimensionality terms in the gluon-meson Lagrangian exist, with more derivative or field terms, the reaction from two gluons to two mesons should dominate during the freeze-out at the boundary of the gluon plasma. Thus we employ this Lagrangian (\ref{lfin}) to evaluate the number of produced pions and kaons per gluon.

\begin{figure}[t!]
\begin{center}
\includegraphics[
width=3.0in
]%
{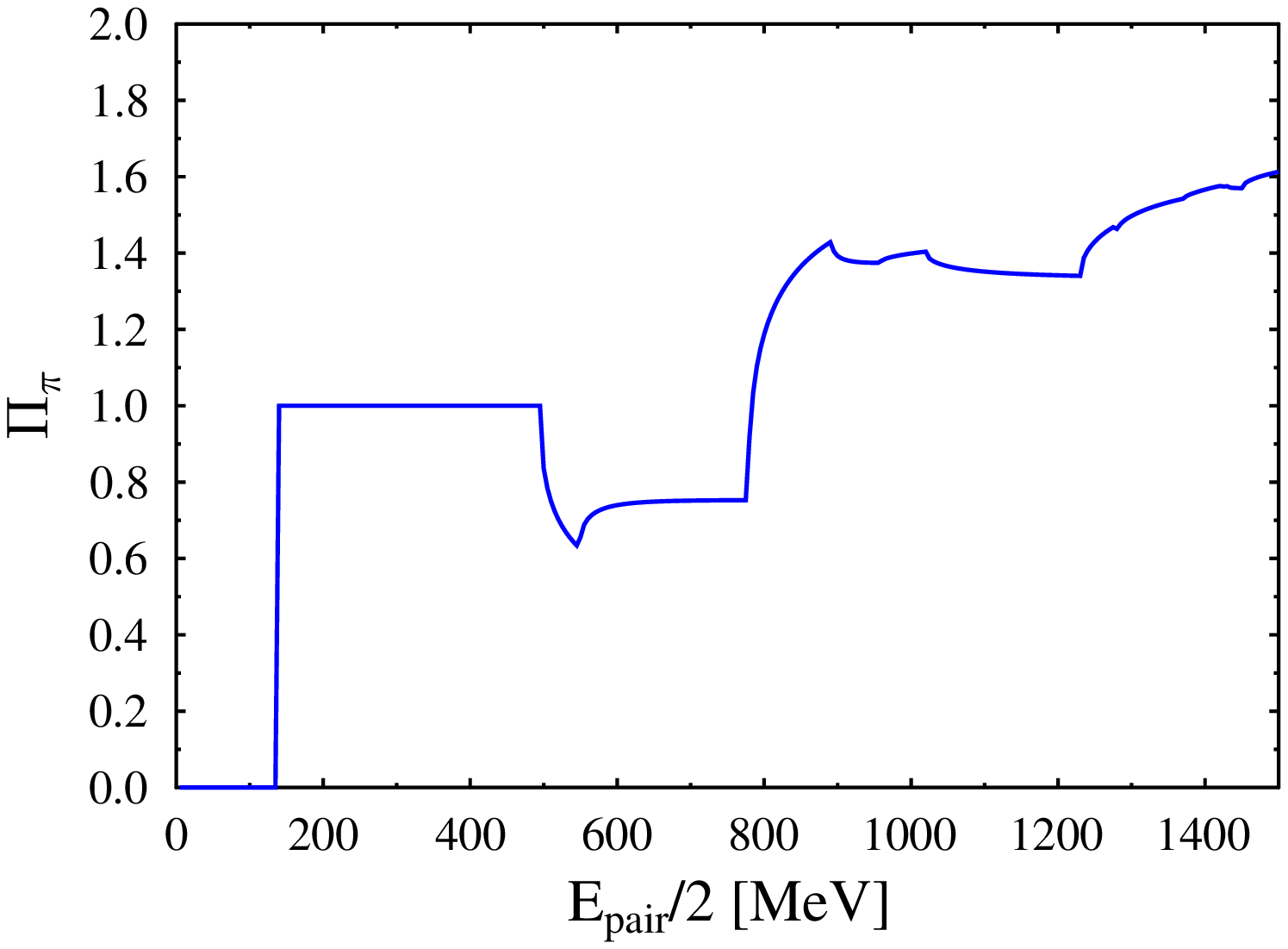}%
\end{center}
\begin{center}
\includegraphics[
width=3.0in
]%
{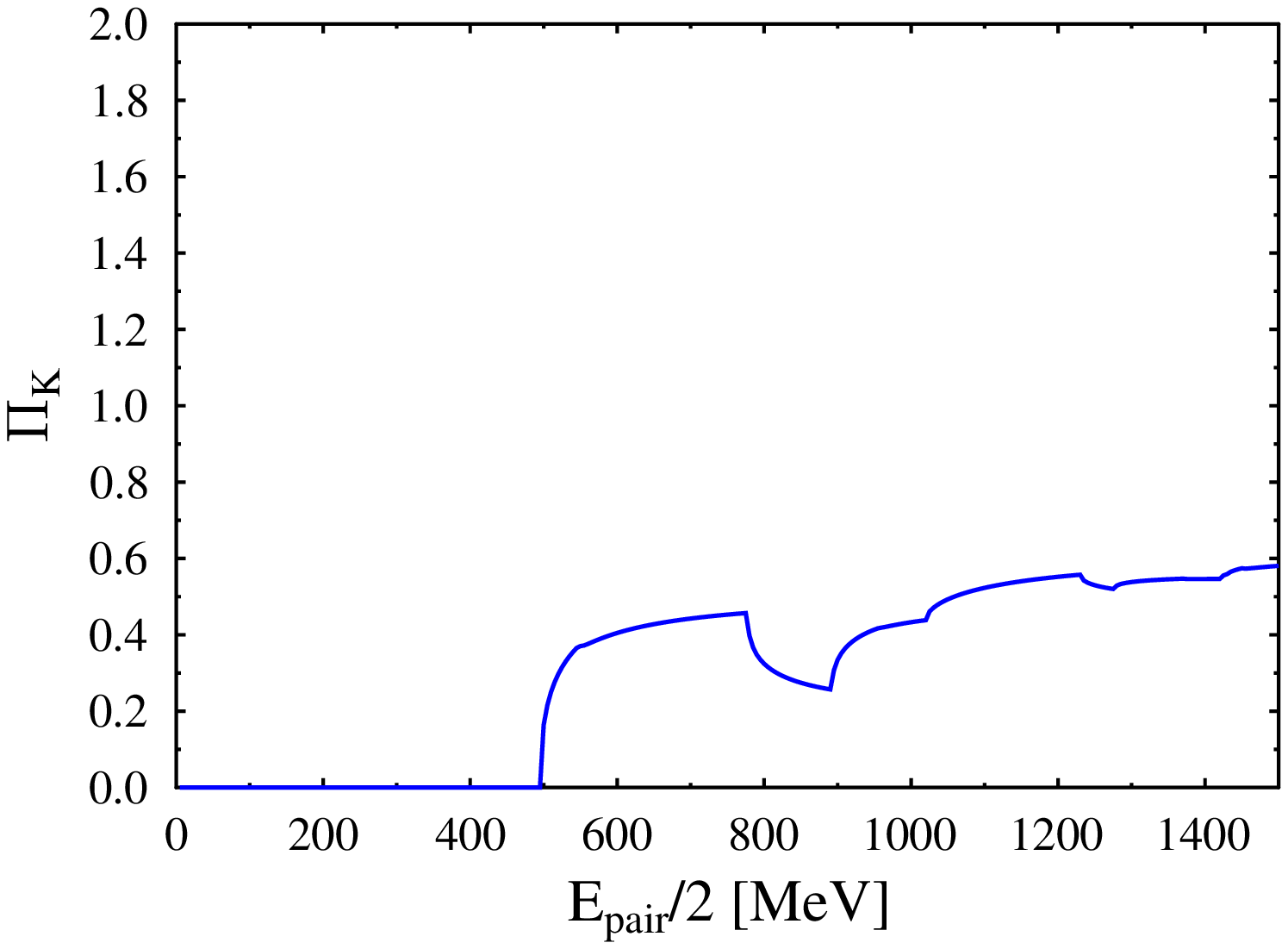}%
\end{center}
\begin{center}
\caption{Number of (top) pions $\Pi_{\pi}(E_{pair})$ and of (bottom) kaons $\Pi_{k}(E_{pair})$ produced per gluon as a
function of the gluon energy in the centre of mass of the gluon pair. Here
$a=b,$ $c=0$ are used. }%
\label{multen}%
\end{center}
\end{figure}

%S
%S
%S
%S
%S
%S
%S
%S
%S
%S
%S
%S

\section{Analytical and numerical results}

In our framework an effective gluon mass $M_{g}$ can be introduced.
Notice the existence of a possible effective gluon mass
\cite{Cornwall:1981zr}, 
or pole in a gluon propagator, already at $T=0$, has been under debate in QCD. While gauge invariance in a perturbative approach rules out a gluon mass and the existing lattice QCD glueballs suggest that the gluon propagator is transverse, nevertheless a running gluon mass does not contradict gauge invariance
\cite{Aguilar:2011xe}, 
and there is also evidence both from Landau Gauge lattice QCD \cite{Oliveira:2010xc} and from the glueball spectrum for a finite pole in the gluon propagator. Most results, several of them from lattice QCD calculations, point to a $T=0$ effective gluon mass or to other possible scales, say an effective dual gluon mass \cite{Cardoso:2010kw}, in the range [0.5, 1.0] GeV. 
A possible gluon mass at finite $T$ is also starting to be investigated in lattice QCD
\cite{Cucchieri:2012nx}.
Moreover in the BAMPS set-up 
a finite Debye mass \cite{Xu:2008av} for the gluon is also considered,
\be
{M_g}^2 = {24 \over \pi} \alpha_s T^2 ,
\ee
which, say at $T=T_c=0.158$ GeV and $\alpha_s \simeq 0.3$ leads to a gluon mass of $M_g \simeq 0.239$  GeV. 

For the following purposes we evaluate the Lorentz-invariant Mandelstam
variable $s$ for a system of two gluons with four-momenta 
\be
{p_{i}}  =(\sqrt{\mathbf{p}_{i}^{2}+M_{g}^{2}},\mathbf{p}_{i}) \ , \ \ i=1,2 \ .
\ee
The Mandelstam variable $s$ leads to the centre of mass energy $E_{pair}$ of the gluon pair,
\bea
s  &  = &({p_{1}}+{p_{2}})^{2}
\\ \non
&  = & 2M_{g}^{2}+2\left(  \sqrt{\mathbf{p}_{1}^{2}+M_{g}^{2}}\sqrt
{\mathbf{p}_{2}^{2}+M_{g}^{2}}-\mathbf{p}_{1}\cdot\mathbf{p}_{2}\right) 
\\ \non
&  = & E_{pair}^{2} \text{ . }%
\eea

\begin{figure}[t!]
\begin{center}
\includegraphics[
width=3.0in
]%
{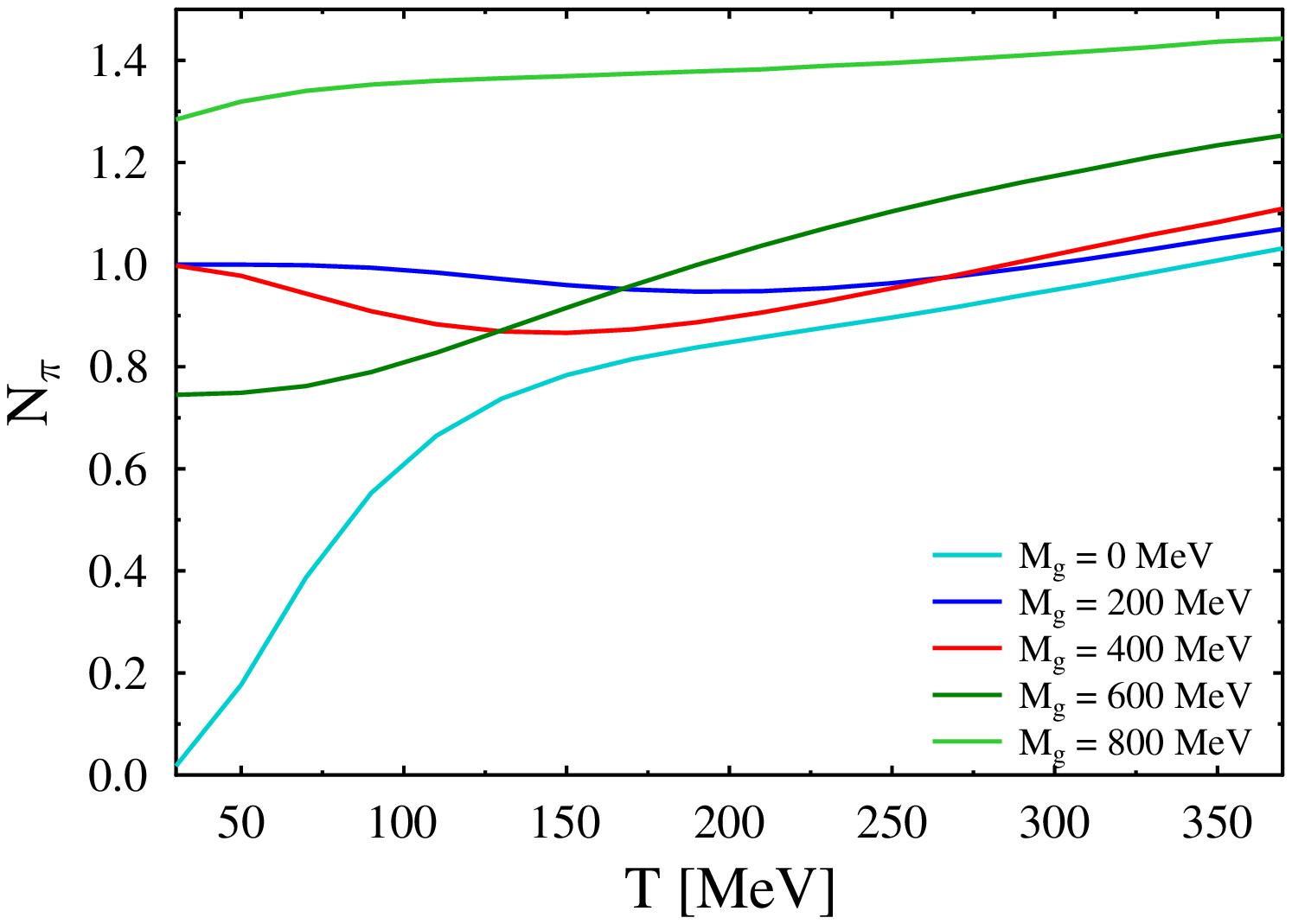}%
\end{center}
\begin{center}
\includegraphics[
width=3.0in
]%
{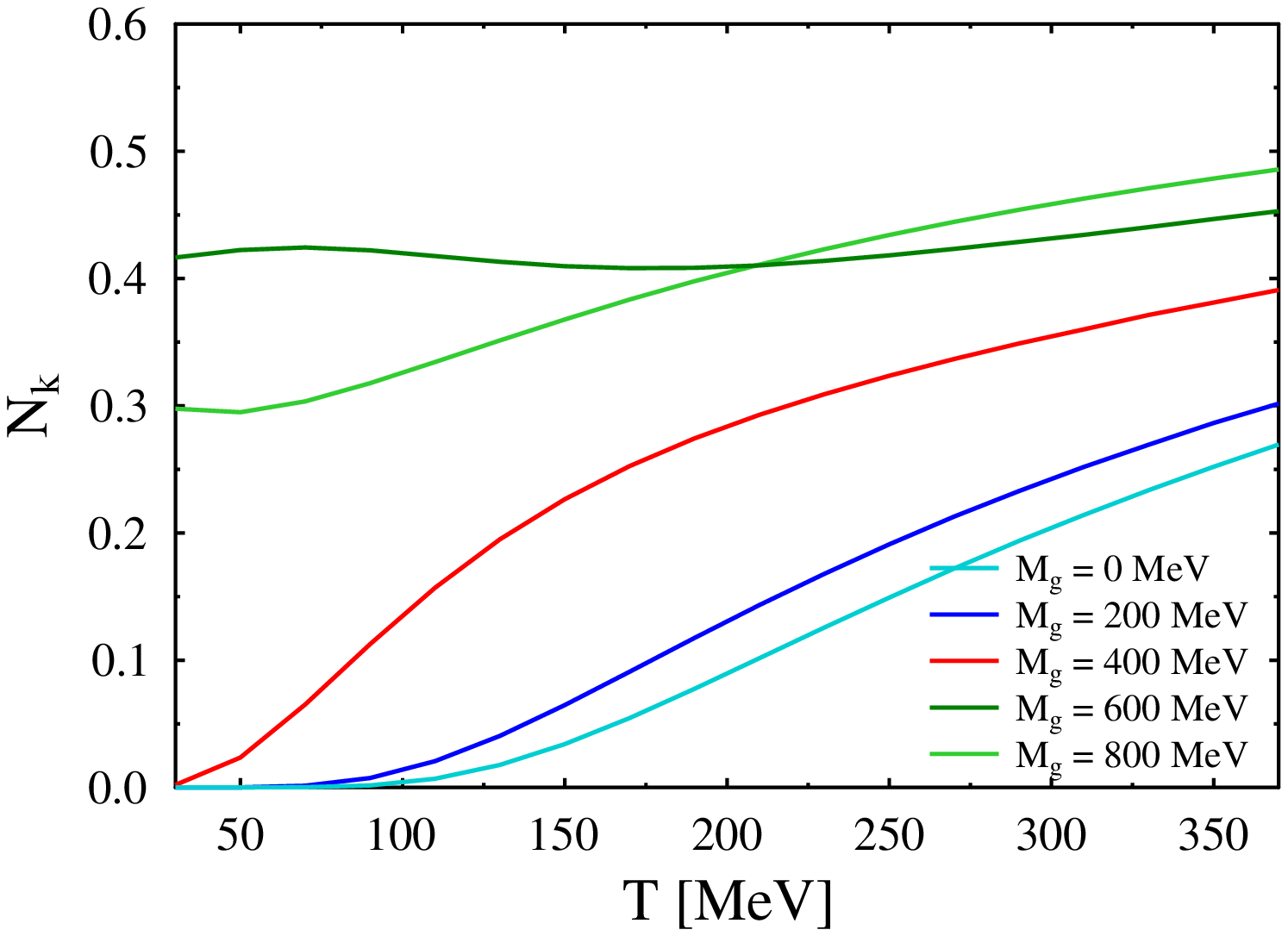}%
\end{center}
\begin{center}
\caption{Number of (top) pions and (bottom) kaons produced per gluon as function of the temperature $T$ for
different gluon masses.}%
\label{multtems}%
\end{center}
\end{figure}

Due to the Lagrangian in Eq. (\ref{lfin}), the two gluons convert into a
meson-pair. Considering that only pions and kaons are regarded as stable, we
must also take into account that the other resonances decay subsequently into
kaons and pions. For instance, each $\sigma$ meson decays into a two-pion
pair, therefore the $\sigma\sigma$ channel results into a final $4\pi$ state.
Similarly, a pair of $\rho$ mesons decays also into four pions. On the
contrary, the vector state $\phi$ decays predominantly into kaons. In Table
\ref{mesons} these conversion factors, expressed as $n^{(\pi)}$ and $n^{(K)}$, are listed for all the relevant mesons used for the calculations.
Moreover, we also include the usual phase space factor $\sqrt{\frac
{E_{pair}^{2}}{4}-{M}_{k}^{2}},$ where $M_{k}$ is the mass of the $k$-th meson
pair, and the corresponding degeneracy spin-isospin factor $g_{k}$. Finally,
we should also take into account the relative strength, which is equal to
$a^{2}$ in the (pseudo)scalar channel, to $b^{2}$ in the (axial-)vector
channel, and to $c^{2}$ in what concerns the scalar channel below 1 GeV. Putting all together, the number of pions and kaons per gluon is
calculated as%
\begin{equation}
\Pi_{i}(E_{pair})={\frac{\sum_{k}\sqrt{\frac{E_{pair}^{2}}{4}-{M}_{k}^{2}%
}\theta(E_{pair}-4{M}_{k})g_{k}{f_{k}}^2n_{k}^{(i)}}{\sum_{k}\sqrt{\frac
{E_{pair}^{2}}{4}-{M}_{k}^{2}}\theta(E_{pair}-4{M}_{k})g_{k} {f_{k}}^2}}\
\end{equation}
where $i=\pi,K.$

The functions $\Pi_{\pi}(E_{pair})$ and $\Pi_{K}(E_{pair})$ as a function of
the gluon energy and as a function of the Boltzmann temperature are both depicted in Fig. \ref{multen}.

\begin{figure}[t!]
\begin{center}
\includegraphics[
width=3.2in
]%
{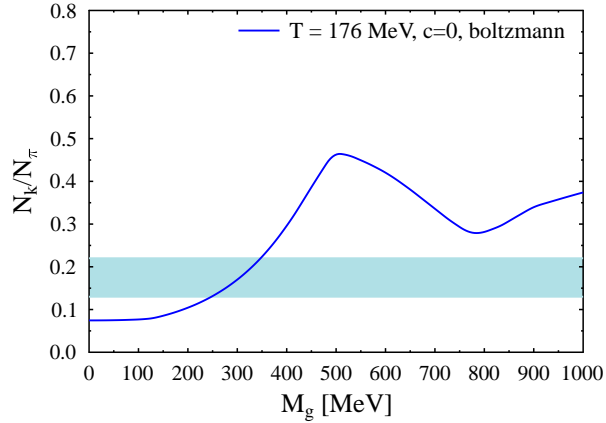}%
\end{center}
\begin{center}
\caption{We show (solid line) our result and (band) the BNL, CERN and FNAL data for the kaon to pion multiplicity ratio as a function of a possible gluon mass, at $T=T_c \simeq 158$ MeV. The measured ratio $N_{K}/N_{\pi} \in [0.13 , 0.22] $ is realized for a gluon mass $M_{g} \in [0.25 , 0.35]$ GeV.}%
\label{ratios}%
\end{center}
\end{figure}

\begin{figure*}
[ptb]
\begin{center}
\includegraphics[
width=2.3in
]%
{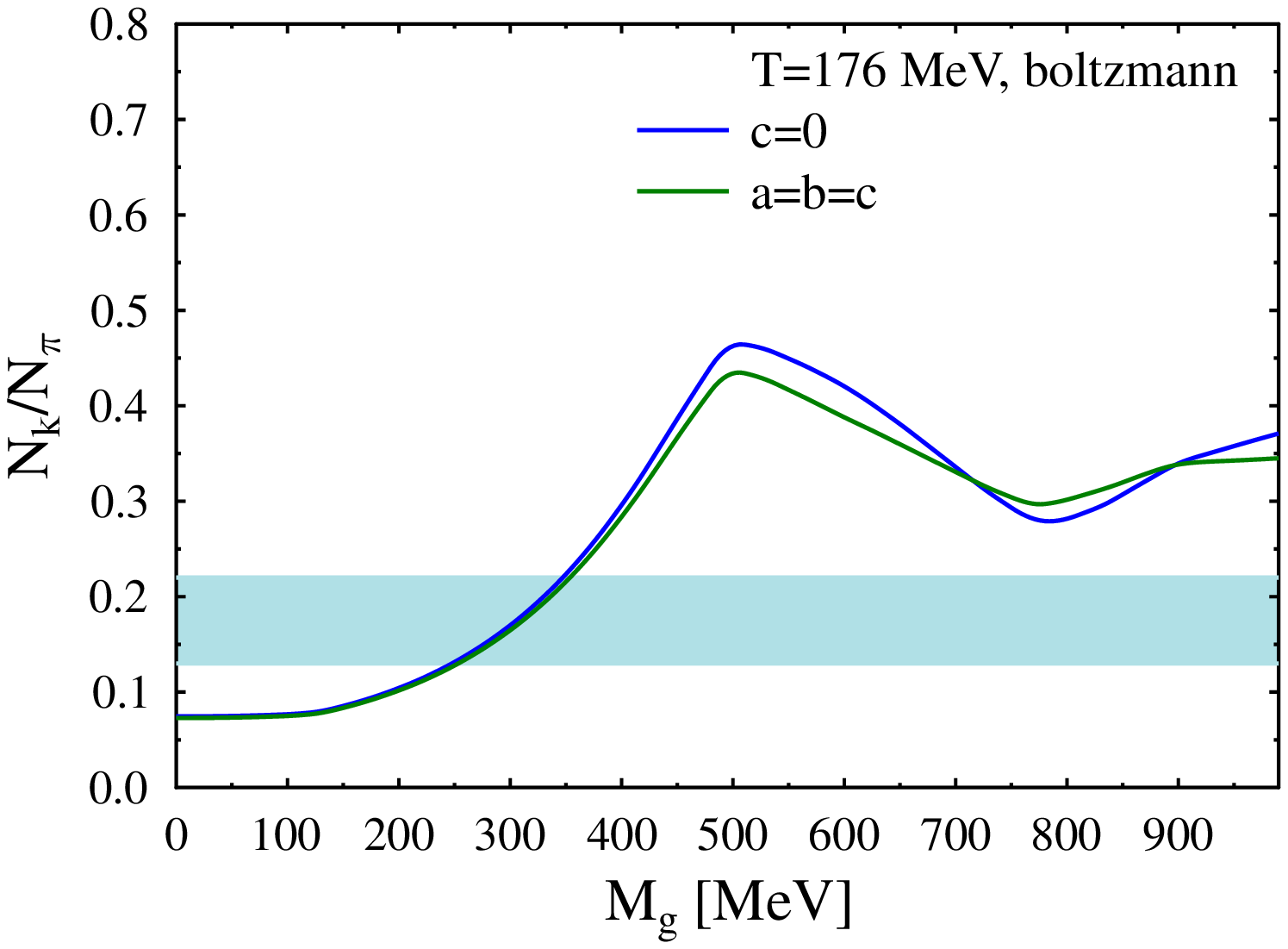}%
\includegraphics[
width=2.3in
]%
{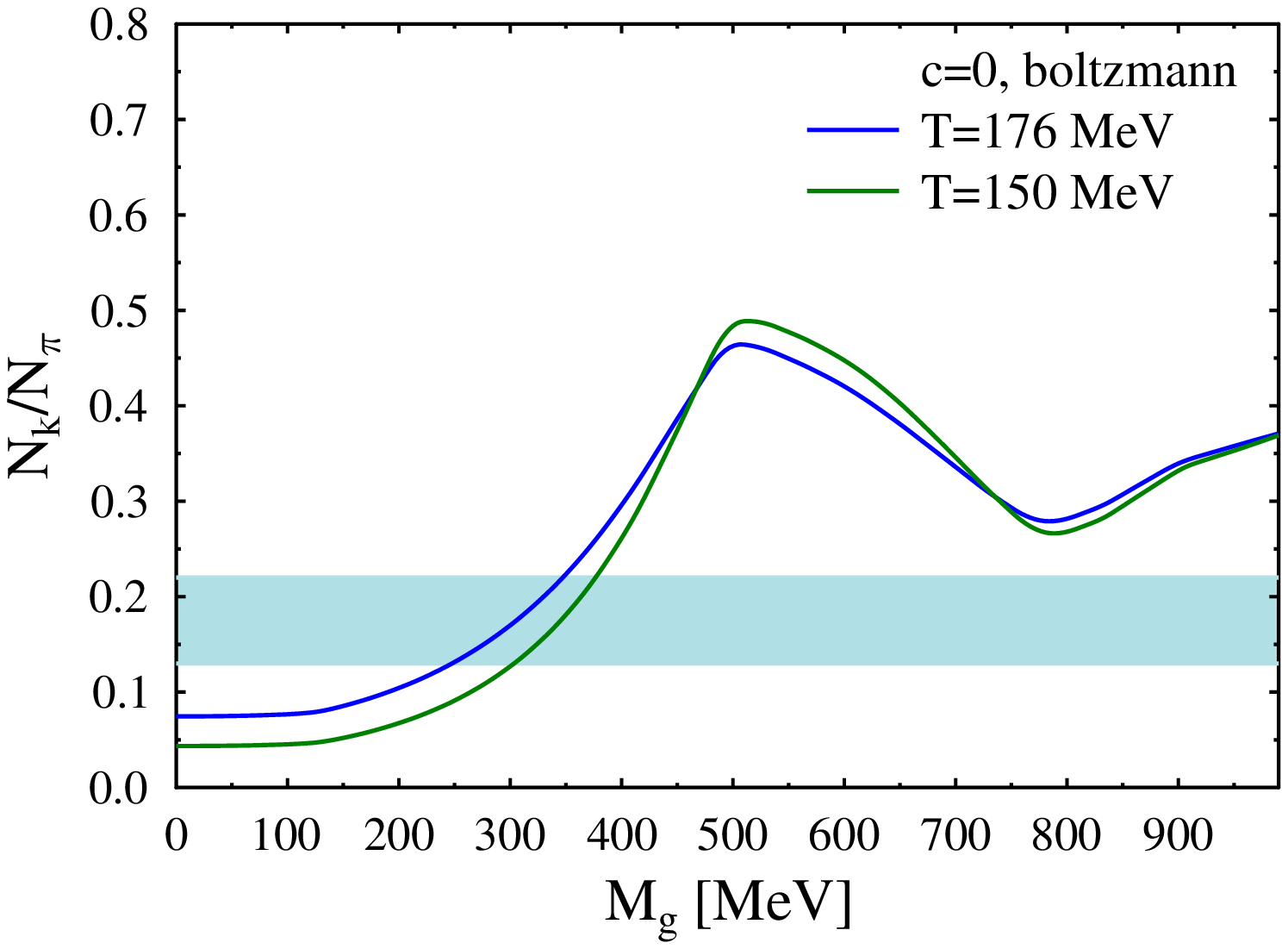}%
\includegraphics[
width=2.3in
]%
{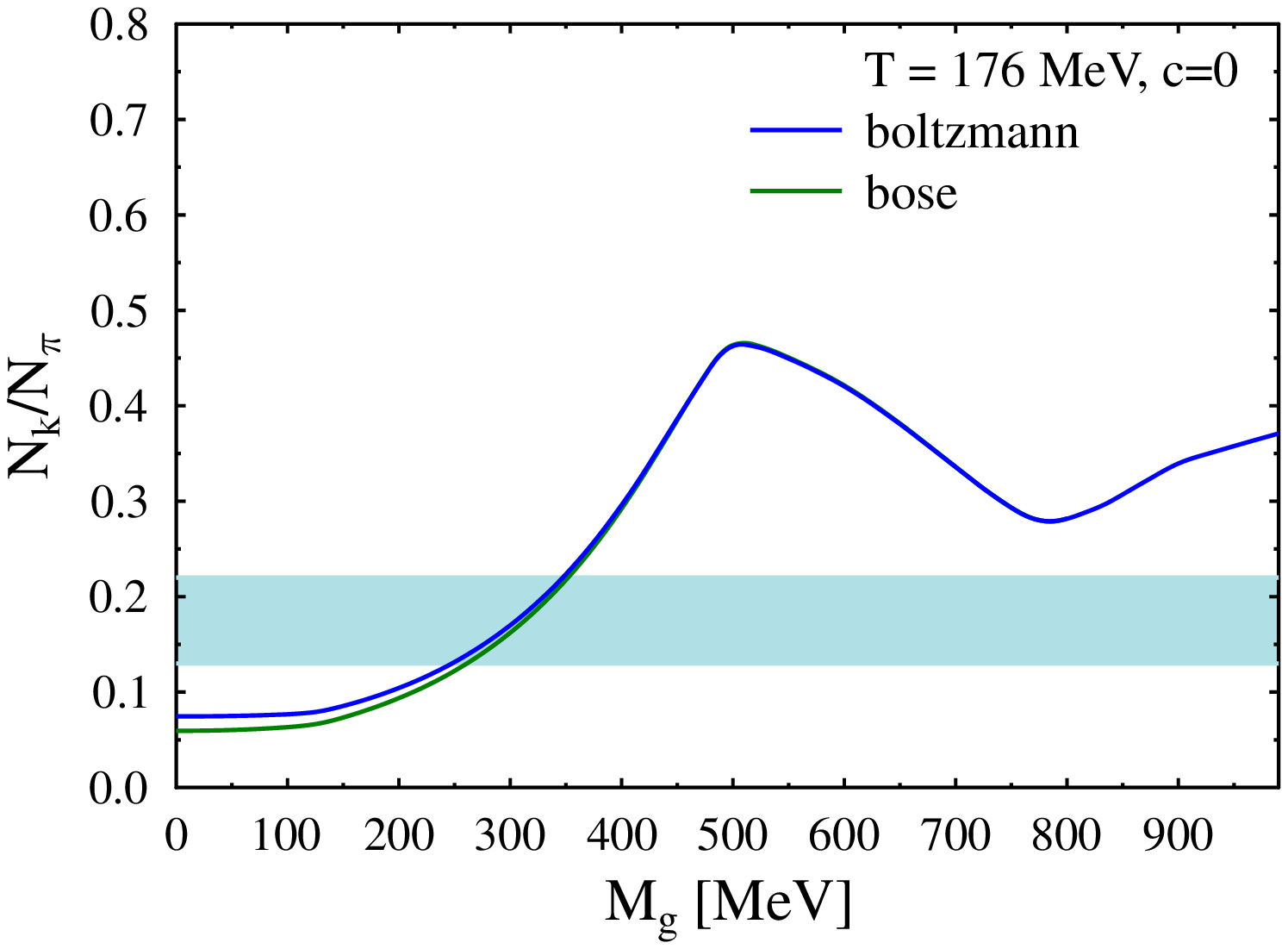}%
\caption{The parameter dependence in the gluon-meson model. We compare the ratio $N_{K}/N_{\pi}$ for different couplings (left, $c=0$ with $c=1$), different temperatures (centre, $T = 0.145$ GeV with $T = 0.170$ GeV) and different statistical distributions (right, Boltzmann and Bose-Einstein).}%
\label{tests}%
\end{center}
\end{figure*}

We are interested in comparing our results with the parametrization of the BAMPS set-up
\cite{Xu:2008av}, since, like our framework, BAMPS also consists of gluons, decaying into pions. In BAMPS, 1.5 to 2.0 pions are produced per gluon pair. This production takes place in different conditions than ours, at non-chemical equilibrium with local and dynamical many-gluon simulations. Nevertheless, if we compare with our approach, we notice we can easily obtain $N_{\pi}\sim 1$, but the increase of $N_{\pi}$ above unity can only be achieved at the price of including a sizeable effective gluon mass.
This confirms the importance of including in our framework an effective gluon mass, simulating a finite non-perturbative scale characteristic of the gluon plasma.

We now evaluate the emitted number of pions $N_{\pi}(T)$ and
kaons $N_{K}(T)$ per gluon as a function of the temperature $T$. We denote the
multiplicities $N_{i}(T)$ by employing a Boltzmann distribution of each gluon,
thus leading to
\begin{equation}
N_{i}(T)=\int d\mathbf{p}_{1}\,d\mathbf{p}_{2}f_{B}(\mathbf{p}_{1}^{2}%
,T)f_{B}(\mathbf{p}_{2}^{2},T)\Pi_{i}(E_{pair})\text{ ,}
\label{3dint}
\end{equation}
where $f_{B}(\mathbf{p}^{2},T)=\mathcal{N}e^{-\sqrt{\mathbf{p}^{2}%
+m_{gluon}^{2}}/T}$ is the normalized Boltzmann distribution. The integral in Eq. (\label{3dint}) can be simplified to a three-dimensional integration, and we compute it with a numerically accurate c++ code. In Fig. \ref{multtems} the functions $N_{\pi}(T)$ and
$N_{K}(T)$ are plotted for different values of the gluon mass $M_{g}$ and for
both choices $a=b,$ $c=0$ and $a=b=c.$ For the intermediate value $M_{g}=400$
MeV we have roughly one pion per each gluon for each $T$, while $N_{K}(T)$ is
a rapidly increasing function with $T$.

In Fig. \ref{ratios} we present the ratio $N_{K}(T)/N_{\pi}(T)$ for the temperature of $T=T_c=0.158$ GeV as a function of
the gluon mass. The critical temperature $T_c$ for the confinement and chiral crossover was measured in lattice QCD to be in the range $T_c \in [0.145,0.165]$ GeV 
\cite{Borsanyi:2010bp,Aoki:2006we,Cheng:2006qk,Detar:2007as}. This was achieved in very precise full lattice QCD simulations with dynamical fermions. 
This temperature range is consistent with the freeze-out temperature of the quark-gluon plasma measured in Heavy ion collisions.
The freeze-out temperature in heavy ion collisions can be
determined from the inverse slope of the hadronic species multiplicity as a function of the transverse momentum.
Recent analysis of heavy ion collisions indicate that the freeze-out temperature is in the range of $[0.150,0.170]$ GeV
with results between 0.150 GeV and 0.160 GeV
\cite{inverseslope150to160} and results between 0.160 GeV and 0.170 GeV \cite{inverseslope165to175,Bratkovskaya}. Both ranges are compatible, and we consider in our computations the mean value of $T=0.158$ GeV.

In Fig. \ref{ratios} we compare the ratio with the experimental data $(N_{K}/N_{\pi})_{\exp} \ \in [1.13, 1.22 ]$ measured by the PHENIX, STAR, BRAHMS, E866 and NA49 collaborations and extrapolated by the UrQMD 2.0, UrQMD 2.1 and HSD transport approaches
\cite{Adcox:2001mf,Bratkovskaya} for the ratio of the pion and kaon multiplicity in the most central collisions. 

Our results point to a solution corresponding to a possible effective gluon mass, in a range of $M_g \in [0.28 , 0.37 ]$ GeV. Remotely, a second less likely mass of circa 0.8 GeV may be possible. We notice that the solution points to a gluon mas at $T_c$ of the order of circa 0.4 of the gluon effective mass of 0.5 to 1.0 GeV at $T=0$ resulting from different gluon calculations. The solution for the gluon mass is also consistent with the Debye mass
\cite{Xu:2008av} of the gluon at finite $T$. 

Finally we test the robustness of our results checking the parameter dependence of the gluon-meson model. In Fig. \ref{tests} we compare the ratio $N_{K}/N_{\pi}$ for different couplings ($c=0$ with $c=1$), different temperatures ($T$ = 0.145 GeV with $T$ = 0.170 GeV) and different statistical distributions (Boltzmann and Bose-Einstein).
All the tests we performed with plausible changes  of our parameters suggest our results are robust. 

The small effect of changing the coupling of gluons to mesons is quite relevant for our results, since the meson properties are not yet established. Strong coupled channel effects, tetraquarks, and glueballs have been shown to affect the meson spectrum and the mesonic couplings. Here we utilize the meson properties listed in the Particle Data Group \cite{PDG}, but other meson data would yield similar results. We utilize the sigma model for the meson production, but other hadronic models would again yield similar results. 

The robustness of our results occurs because a gluon in the freeze-out of the plasma has a low energy ($T \simeq$ 160 MeV and an effective mass of $M_g \simeq$ 320 MeV), much lower than the energy of a gluon in any glueball typical of lattice QCD simulations or of  constituent gluon model estimations.
Thus our results do not really depend on the details of the meson
spectrum and of the meson couplings above that low energy, and escape the problem of understanding  higher energy reactions such as the glueball decays.

%S
%S
%S
%S
%S
%S
%S
%S
%S
%S
%S
%S

\section{Conclusions}

In this work we develop an effective Lagrangian which connects gluons to mesons. We then utilize this approach to calculate the emitted number of pions and kaons per gluon out of a gluon gas at temperature $T$. We assume the gluons are in a thermal bath at $T=T_c$ whereas the mesons are produced at vanishing temperature.

The fact that our effective Lagrangian consists of bosonic degrees of freedom only might be also interesting for lattice QCD applications. In fact, the Grassmann variable nature of the quarks makes them technically much more difficult to address (see for instance Ref. \cite{Gattringer:2010zz} and refs. therein) than the bosonic gluons and mesons.

Our approach represents an attempt to link directly and in an understandable way the gluon-dominated physics of the quark gluon plasma, as suggested by the colour glass condensate and BAMPS approaches, to the light hadrons in the final stage. 

Further developments of our approach could include meson mass modifications in the medium, effects of the freeze-out boundary, different meson-gluon couplings, missing resonances such as the tensor mesons, final state interactions among mesons, and, last but not least, glueball fields, which directly couple to gluonic fields.
Nevertheless, all the tests we performed with plausible changes  of our parameters suggest that our results are stable. Thus we regard our effective Lagrangian as a pilot study toward a better understanding of the rich physics of QCD and of Heavy Ion Collisions.

Interestingly, to reproduce the experimental results of the PHENIX, STAR, BRAHMS, E866 and NA49 collaborations for the pion and kaon multiplicities, we have to include a finite scale for the gluon energy at $T=T_c$. 
Our results point to a possible effective gluon mass, in a range of $M_g$ between 0.28 and 0.37 GeV if we consider the uncertainty on the rapidity ratio, or of $M_g \in [0.25 , 0.39 ]$ if we also consider the uncertainty on the freeze-out temperature. We notice that this solution points to a gluon mass at $T_c$ of the order of circa 0.4 and 1.0 of the gluon effective mass of 0.5 to 1.0 GeV at $T=0$ resulting from different gluon calculations. An $M_g \in [0.25 , 0.39 ]$ is also close to the Debye gluon mass at $T=T_c$.
Notice this solution corresponds to an absolute pion multiplicity of circa 0.9 pions per gluon.  

If the gluon mass is related to confinement, say as in superconductors, our result is consistent with the QCD order parameters at $T_c$. Both for a first order phase transition
where the order parameter is discontinuous
 (found in quenched lattice QCD \cite{Cardoso:2011hh}) and in a crossover where the order parameter remains finite (found in lattice QCD with dynamical fermions \cite{Borsanyi:2010bp}), the scale of confinement should not simply vanish at $T=T_c$. Here we present an evidence, based in the experimental Heavy Ion data and in our simple and robust gluon-meson model that a finite scale of 0.25 to 0.39 GeV exists in the gluon sector of QCD at $T=T_c$.

\acknowledgments{ We thank discussions with Marlene Nahrgang, Magdalena Malek, Edmond Iancu, George Rupp, Andrey El, Ianni Bouras, Giorgio Torrieri and Marcus Bleicher. This work was partly funded by the CRUP/DAAD Luso-German exchange AI-A-2010-10.}

%B
%B
%B
%B
%B
%B
%B
%B
%B
%B
%B
%B

\end{document}